\documentclass
[superscriptaddress,secnumarabic,amssymb,amsmath,nobibnotes,aps,prd,showkeys,showpacs,nofootinbib,onecolumn]{revtex4}%
\usepackage{graphicx}
\usepackage{epsf}
\usepackage{bm}
\usepackage{amsmath}
\usepackage{amsfonts}
\usepackage{amssymb}
\usepackage{epstopdf}%
\providecommand{\U}[1]{\protect\rule{.1in}{.1in}}

\newcommand{\be}{\begin{equation}}
\newcommand{\ee}{\end{equation}}

\newcommand{\mincir}{\raise
-3.truept\hbox{\rlap{\hbox{$\sim$}}\raise4.truept\hbox{$<$}\ }}
\newcommand{\magcir}{\raise
-3.truept\hbox{\rlap{\hbox{$\sim$}}\raise4.truept\hbox{$>$}\ }}

\ifx\pdfoutput\relax\let\pdfoutput=\undefined\fi
\newcount\msipdfoutput
\ifx\pdfoutput\undefined\else
\ifcase\pdfoutput\else
\ifx\paperwidth\undefined\else
\ifdim\paperheight=0pt\relax\else\pdfpageheight\paperheight\fi
\ifdim\paperwidth=0pt\relax\else\pdfpagewidth\paperwidth\fi
\fi\fi\fi
\begin{document}
\title{Dynamical symmetries in Brans-Dicke cosmology}
\author{G. Papagiannopoulos}
\email{yiannis.papayiannopoulos@gmail.com}
\affiliation{Faculty of Physics, Department of Astronomy-Astrophysics-Mechanics University
of Athens, Panepistemiopolis, Athens 157 83, Greece}
\author{John D. Barrow}
\email{jdb34@hermes.cam.ac.uk}
\affiliation{DAMTP, Centre for Mathematical Sciences, University of Cambridge, Wilberforce
Rd., Cambridge CB3 0WA, U.K.}
\author{S. Basilakos}
\email{svasil@academyofathens.gr}
\affiliation{Academy of Athens, Research Center for Astronomy and Applied Mathematics,
Soranou Efesiou 4, 11527, Athens, Greece}
\author{A. Giacomini}
\email{alexgiacomini@uach.cl}
\affiliation{Instituto de Ciencias F\'{\i}sicas y Matem\'{a}ticas, Universidad Austral de
Chile, Valdivia, Chile}
\author{A. Paliathanasis}
\email{anpaliat@phys.uoa.gr}
\affiliation{Instituto de Ciencias F\'{\i}sicas y Matem\'{a}ticas, Universidad Austral de
Chile, Valdivia, Chile}

\begin{abstract}
In the context of generalised Brans-Dicke cosmology we use the Killing tensors
of the minisuperspace in order to determine the unspecified potential of a
scalar-tensor gravity theory. Specifically, based on the existence of contact
symmetries of the field equations, we find four types of potentials which
provide exactly integrable dynamical systems. We investigate the dynamical
properties of these potentials by using a critical point analysis and we find
solutions which lead to cosmic acceleration and under specific conditions we
can have de-Sitter points as stable late-time attractors.

\end{abstract}
\maketitle
\date{\today}

\section{Introduction}

The analysis of the recent detailed cosmological data
\cite{Teg,Kowal,Komatsu,suzuki11,Ade15} has posed new problems for modern
cosmology. One of the most fundamental is the nature of the so called 'dark
energy' driving the late-time acceleration of the universe. The possible
answers follow two different approaches. One of these proposes to modify the
theory of gravity with new terms in the Einstein-Hilbert action which create
anti-gravitating effects that mimic the presence of an inflationary
self-interacting scalar field, see \cite{Aref1,Aref2,Aref3,Aref4,Aref5,Aref6}
and references therein. Alternatively, we can introduce an \textquotedblleft
exotic\textquotedblright\ matter source directly into the arena of general
relativity in order to explain the late-time acceleration
\cite{Ar1,Ar2,Ar3,Ar4,Ar5,Ar6,Ar7,Ar8}. However, in both of these approaches,
the complexity of the field equations is increased considerably and
qualitative methods in the theory of differential equations are needed to
supplement the search for exact solutions of the cosmological equations.

The method of group invariant transformations is a powerful tool for the
derivation of conservation laws of differential equations. In gravitational
theories, the so called point symmetries (specifically, the ones which satisfy
Noether's theorem) have been used to derive new integrable systems and then
new exact and analytical solutions (see
\cite{ns0,ns2,ns2a,ns3,ns4,ns4a,AnScT,ns5,ns7,anb,belin01,hao,biswas} and
references therein). However, another family of symmetries are the so-called
'contact symmetries'. These are the generators of the infinitesimal
Lie-B\"{a}cklund transformations which are linear in the first derivatives of
the unknown functions. These symmetries provide quadratic conservation laws,
in contrary to 'point symmetries' which provide linear conservation laws. The
application of contact symmetries in scalar field cosmology and in the case of
$f\left(  R\right)  $-gravity can be found in \cite{kt1,kt2}. Of course, there
are other methods which can be used to determine conservation laws and study
integrable cosmological modes. Some of them are related with the search of
group invariant transformations; for instance, see
\cite{ref01,ref02,ref03,ref04,ref06,ref07,ref08,ref09,ref10,ref11,ref12,refAA1,refAA2}
and references therein.

In this paper we explore the application of the Killing tensors of the
minisuperspace which generate contact symmetries (quadratic conservation laws)
for the field equations in Brans-Dicke gravitational theory. This theory was
introduced for the first time in \cite{Jord, BD} and used by Brans and Dicke
to construct a theory of gravity which embodied Mach's Principle. Indeed, in
this theory the gravitational field is described not only by the metric tensor
but also by a non-minimally coupled scalar field, which plays an important
role in the description of the early universe as well as in the era of dark
energy \cite{ssen}. After the discovery of cosmic acceleration a particular
attention has been paid on the Brans-Dicke theory by several authors
\cite{Boi} (for review see \cite{Clif12} and references therein). In
particular the dynamical analysis of the Brans-Dicke model has revealed, under
specific conditions, two critical points \cite{DEAmendola}. The first one is
related with a de-Sitter point which implies an accelerated phase of the
universe, while the second critical point corresponds to the radiation
dominated era. It is well known that the Brans-Dicke theory is characterized
by the so-called Brans-Dicke parameter $\omega_{BD}$ which provides an
effective Newton's parameter and it determines the strength of the coupling
between the scalar field and the matter sources. Notice, that large values of
$\omega_{BD}$ mean a significant contribution from the tensor part, while
small values of $\omega_{BD}$ imply that the contribution from the scalar
field is significant. While it is expected that in the limit $\omega
_{BD}\rightarrow\infty$ one would recover General Relativity, it was however
noticed that in general the latter is not true, which means that the two
theories- as shown by \cite{omegaBDGR}- are fundamentally different. In this
context, using Solar System data it has been \cite{BER} that the Brans-Dicke
parameter is constrained to be $\omega_{BD}>4\times10^{4}$ at 2$\sigma$ level
(see also \cite{BDconstraints}). Other bounds have been found by \cite{Lid}
and \cite{Chen} using the power spectrum of galaxies.  Lastly, it is
interesting to mention that under specific conditions Brans-Dicke theory is
related to another very popular generalization of general relativity, namely
$f(R)$ gravity. Indeed, in the case of $\omega_{BD}=0$ Brans-Dicke theory
becomes equivalent to $f(R)$ gravity in the metric formalism, whereas if
$\omega_{BD}=-\frac{3}{2}$ it becomes equivalent to $f(R)$ gravity in the
Palatini formalism \cite{Sotiriou}. We also note that the low-energy limit of
bosonic string theory is described by a Brans-Dicke theory with $\omega
_{BD}=-1$ \cite{string}. An interesting explanation of the possible origin of
the scalar field comes from Kaluza-Klein compactification which gives
Brans-Dicke theory with $\omega_{BD}=-\frac{d-1}{d}$, where $d$ is the number
of extra space dimensions \cite{cho}. Recently, Brans-Dicke theory has also
attracted attention in cosmology in the context of inflationary scenarios (see
\cite{faraonibook} and references therein). The aim of this paper is to study
the cosmological dynamics for a flat FLRW spacetime where the scalar field
inherits the symmetries of the spacetime. Furthermore, in order to study the
basic properties of the Brans-Dicke theories, which follow from the
application of group invariant transformations, we study the critical points
of the field equations for an arbitrary potential in dimensionless variables.
\newline

The plan of the paper is as follows. In Section \ref{section2} we give the
main features of the Brans-Dicke cosmological model in a FLRW spacetime and
present the field equations in the minisuperspace approach. In Section
\ref{section3} we search for contact symmetries generated by the Killing
tensors of the minisuperspace. In particular, we find four different families
of potentials (ie theories) in which the field equations admit quadratic
conservation laws. The field equations form dynamical systems which are
Liouville integrable; that is, they can be solved by quadratures. In Section
\ref{section4} we investigate the cosmological evolution for the vacuum
Brans-Dicke theory with an arbitrary potential for the scalar field. We see
that the cosmological models derived from the application of Killing tensors
accommodate cosmic acceleration universe and de Sitter phases. Finally, we
draw our conclusions in Section \ref{conclusions}.

\section{Brans-Dicke theory}

\label{section2}

The action which describes the gravitational field equations and satisfies
Mach's principle in some appropriate form was first introduced by Brans and
Dicke in \cite{bdpaper}. The action integral is defined as follows%
\begin{equation}
S=\int dx^{4}\sqrt{-g}\left[  \frac{1}{2}\phi R-\frac{1}{2}\frac{\omega_{BD}%
}{\phi}g^{\mu\nu}\phi_{;\mu}\phi_{;\nu}-V\left(  \phi\right)  \right]  ,
\label{bd.01}%
\end{equation}
where $\phi~$is the Brans-Dicke scalar field, $\omega_{BD}$ is the Brans-Dicke
parameter and $R$ is the Ricci Scalar of the underlying spacetime. The
gravitational field equations follow from the variation of the action
(\ref{bd.01}) with respect to the metric tensor $g_{\mu\nu}$, while the
equation of motion for the field $\phi$ follows under the variation with
respect to $\phi$. The original Brans-Dicke theory took the potential
$V(\phi)$ to be zero but we shall retain it in what follows. Equivalently, it
is possible to set $V(\phi)=0$ but allow the coupling $\omega_{BD}(\phi)$ to
be non-constant.

Varying the action (\ref{bd.01}) with respect to the metric tensor, we arrive
at the field equations,%
\begin{equation}
\phi G_{\mu\nu}=\frac{\omega_{BD}}{\phi}\left(  \phi_{;\mu}\phi_{;\nu}%
-\frac{1}{2}g_{\mu\nu}g^{\kappa\lambda}\phi_{;\kappa}\phi_{;\lambda}\right)
-g_{\mu\nu}V\left(  \phi\right)  -\left(  g_{\mu\nu}g^{\kappa\lambda}%
\phi_{;\kappa\lambda}-\phi_{;\mu}\phi_{;\nu}\right)  , \label{bd.02}%
\end{equation}
in which $G_{\mu\nu}$ is the Einstein tensor. Furthermore, varying equation
(\ref{bd.01}) with respect to the field $\phi~$we obtain the modified
\textquotedblleft Klein-Gordon\textquotedblright\ equation%
\begin{equation}
g^{\mu\nu}\phi_{;\mu\nu}-\frac{1}{2\phi}g^{\mu\nu}\phi_{;\mu}\phi_{;\nu}%
+\frac{\phi}{2\omega_{BD}}\left(  R-2V_{,\phi}\right)  =0. \label{bd.04}%
\end{equation}

Another way to write the gravitational field equations (\ref{bd.02}) is via
the following expression
\begin{equation}
G_{\mu\nu}=\kappa_{eff}\left(  \phi\right)  \left(  T_{\mu\nu}^{\left(
\phi\right)  }+T_{\mu\nu}^{\left(  m\right)  }\right)  , \label{bd.04a}%
\end{equation}
where $\kappa_{eff}\left(  \phi\right)  =\kappa\phi^{-1}$, $\kappa$ is
Einstein's constant, $T_{\mu\nu}^{\left(  \phi\right)  }$ is the effective
energy-momentum tensor for the scalar field $\phi$, that is,%
\begin{equation}
\kappa T_{\mu\nu}^{\left(  \phi\right)  }=\frac{\omega_{BD}}{\phi}\left(
\phi_{;\mu}\phi_{;\nu}-\frac{1}{2}g_{\mu\nu}g^{\kappa\lambda}\phi_{;\kappa
}\phi_{;\lambda}\right)  -g_{\mu\nu}V\left(  \phi\right)  -\left(  g_{\mu\nu
}g^{\kappa\lambda}\phi_{;\kappa\lambda}-\phi_{;\mu}\phi_{;\nu}\right)  ,
\label{bd.04b}%
\end{equation}
and $T_{\mu\nu}^{\left(  m\right)  }$ is the energy momentum tensor of the
matter source, for example a perfect fluid. If we assume that $T_{\mu\nu
}^{\left(  m\right)  }$ is minimally coupled to the scalar field then we get
the usual conservation law $T_{~\ \ ~~~~~~;\nu}^{\left(  m\right)  \mu\nu}=0$.

Here, we assume that the universe is described by the spatially-flat FLRW
spacetime metric with line element%
\begin{equation}
ds^{2}=-dt^{2}+a^{2}\left(  t\right)  \left(  dx^{2}+dy^{2}+dz^{2}\right)  ,
\label{bd.05}%
\end{equation}
from which we define the Ricci scalar\ $R=6\left[  \frac{\ddot{a}}{a}+\left(
\frac{\dot{a}}{a}\right)  ^{2}\right]  $. Consequently, if we assume that the
isometries of (\ref{bd.05}) are symmetries of (\ref{bd.02}) then it follows
that $\phi\equiv\phi\left(  t\right)  $. \ Hence, the gravitational field
equations (\ref{bd.04a}) become \cite{faraonibook}%
\begin{equation}
3H^{2}=\frac{\omega_{BD}}{2}\left(  \frac{\dot{\phi}}{\phi}\right)  ^{2}%
+\frac{V\left(  \phi\right)  }{\phi}-3H\frac{\dot{\phi}}{\phi}+\frac{\kappa
}{\phi}\rho_{m0}a^{-3}, \label{bd.001}%
\end{equation}

\begin{equation}
\dot{H}=\frac{-8\pi}{(2\omega_{BD}+3)\phi}(\omega_{BD}+2)\rho_{m0}%
-\frac{\omega_{BD}}{2}(\frac{\dot{\phi}}{\phi})^{2}+2H(\frac{\dot{\phi}}{\phi
})+\frac{1}{2(2\omega_{BD}+3)\phi}(\phi\frac{dV}{d\phi}-2V), \label{bd.002}%
\end{equation}
where $H=\frac{\dot{a}}{a}.~$\ Overdots indicate differentiation with respect
to the comoving proper time $t$,~and we have assumed that the matter source
$T_{\mu\nu}^{\left(  m\right)  }$ corresponds to dust ($w_{m}=\frac{p_{m}%
}{\rho_{m}}=0$) and describes the dark matter component of the universe.
Notice that the Brans-Dicke potential $V(\phi)$ behaves like a variable
cosmological term, $\Lambda(\phi)=V(\phi)/\phi$, \cite{bmaeda}. Finally, the
modified \textquotedblleft Klein-Gordon\textquotedblright\ equation
(\ref{bd.04}) takes the form:%
\begin{equation}
\ddot{\phi}+3H\dot{\phi}=\frac{8\pi\rho_{m0}-\phi\frac{dV}{d\phi}+2V}%
{2\omega_{BD}+3}. \label{bd.003}%
\end{equation}

\subsection{Minisuperspace approach}

We use the fact that the gravitational field equations can be described by the
basic tools of analytical mechanics. Specifically, the field equations
(\ref{bd.002}), (\ref{bd.003}) follow from the application of the
Euler-Lagrange vector to the Lagrangian function
\begin{equation}
\mathcal{L}\left(  a,\dot{a},\phi,\dot{\phi}\right)  =-3a\phi\dot{a}%
^{2}-3a^{2}\dot{a}\dot{\phi}+\frac{\omega_{BD}}{2\phi}a^{3}\dot{\phi}%
^{2}-a^{3}V\left(  \phi\right)  , \label{bd.07}%
\end{equation}
while eq.(\ref{bd.001}) can be seen as the conservation law of
energy~$\mathcal{E}$. Recall that eq.(\ref{bd.07}) is autonomous and that
$\partial_{t}$ is the trivial Noetherian symmetry. One can easily show that
$\mathcal{E}$ is related with the energy density of the dust fluid, namely
that $\mathcal{E=}$ $\kappa\rho_{m0}$.

For $\omega_{BD}\neq-\frac{3}{2}$, eqn. (\ref{bd.07}) boils down to a
point-like Lagrangian which describes the motion of a particle in the
two-dimensional space
\begin{equation}
ds_{\left(  \gamma\right)  }^{2}=-6a\phi da^{2}-6a^{2}dad\phi+\frac
{\omega_{BD}}{\phi}a^{3}d\phi^{2}, \label{bd.08}%
\end{equation}
under the influence of the potential $V_{eff}\left(  a,\phi\right)
=a^{3}V\left(  \phi\right)  $. Despite the fact that Brans-Dicke is a
second-order theory in the case of $\omega_{BD}=0$, it reduces to the O'Hanlon
theory \cite{Hanlon} which is equivalent to a fourth-order theory, namely the
$f\left(  R\right)  $-gravity in the metric formalism \cite{Sotiriou}. \ 

On the other hand, the limit $\omega_{BD}=-\frac{3}{2}$, (not used here),
corresponds to a second-order theory and the number of degrees of freedom is
the same as for general relativity. This means that the scalar field is not a
real degree of freedom and \ can be eliminated from the dynamics under a
specific \textquotedblleft coordinate transformation\textquotedblright.\ In
this context Brans-Dicke theory is equivalent to a second-order theory:
$F\left(  \mathcal{R}\right)  $-gravity in the affine formalism
\cite{Sotiriou}.

Of course, for a constant field $\phi$, we have either $f_{,RR}=0$ or
$F_{,\mathcal{RR}}=0$, which means that general relativity is recovered.

Now we continue our analysis with the determination of specific forms for the
potential $V\left(  \phi\right)  $, with which the gravitational field
equations admit a second quadratic conservation law in the momentum and form a
Liouville integrable system. We follow the method that applied in the case of
a minimally coupled scalar-field \cite{kt1} and in $f\left(  R\right)
$-gravity \cite{kt2}. Hence, we search for the Killing tensors of the
minisuperspace (\ref{bd.08}) which define contact symmetries for the field
equations and from the second Noether's theorem \cite{sarlet,Kalotas}.

Here, it is important to note that, for the minisuperspace (\ref{bd.08}), the
Ricci Scalar is zero, $R_{\left(  \gamma\right)  }=0$, and since the dimension
of this space is two, the minisuperspace is a flat space. This means that the
minisuperspace admits the three-dimensional isometry group $E_{2}$, while it
admits five independent Killing tensors.

\section{Killing tensors and analytical solutions}

\label{section3}

We start by discussing the relations among the Killing tensors and
conservation laws. Let $\mathcal{H}=\frac{1}{2}\gamma^{ij}p_{i}p_{j}+V$ be the
Hamiltonian function which defines the dynamical system.\ Therefore, the
quadratic function
\begin{equation}
\mathcal{I}=K^{ij}p_{i}p_{j}+\sigma, \label{bd.09}%
\end{equation}
is conserved for the dynamical system with Hamiltonian $\mathcal{H}$, if
$\frac{d}{dt}\mathcal{I}=0;$ or equivalently, if $\{\mathcal{I},\mathcal{H}%
\}=0$. In the case where $K^{ij}$ is a Killing tensor of the metric
$\gamma^{ij}$, that is\footnote{Where $\left[  ,\right]  _{SN},$ denotes the
\ Schouten--Nijenhuis bracket.}, $\left[  \mathbf{K},\mathbf{\gamma}\right]
_{SN}=K_{\left(  ij;k\right)  }=0$, and the following condition holds
\cite{Kalotas}
\begin{equation}
K_{(i}^{~~~~j}V_{,j)}+\sigma_{,i}=0. \label{bd.10}%
\end{equation}

The corresponding symmetry which provides the conservation law (\ref{bd.09})
is called a contact symmetry and it is given as $X=K^{ij}p_{j}\partial_{i}$.
Furthermore, \ the quantity $\sigma$ is the boundary term which is introduced
to allow for the infinitesimal changes in the value of the action integral
produced by the infinitesimal change in the boundary of the domain caused by
the infinitesimal transformation of the variables in the action integral with
generator $X$.

A generalization of the above result has been proved recently in \cite{dim1}
for constrained Lagrangians. However, here we would like to remain in the
framework where the Brans-Dicke field is minimally coupled to the matter
source, so the results of \cite{Kalotas} are applied.

Below, we omit the calculations and we present the results. We find that for
the gravitational field equations (\ref{bd.001})-(\ref{bd.003}), except the
generic constant symmetry $X=\gamma^{ij}p_{j}\partial_{i}$, with corresponding
conservation law the Hamiltonian function, there are four specific choices of
potential function, $V\left(  \phi\right)  $, for which a quadratic in the
momentum conservation law exists:

\subsection{Potential A}

For a Brans-Dicke potential of the functional form
\begin{equation}
V_{A}\left(  \phi\right)  =V_{1}\phi+V_{2}\phi^{-6\omega_{BD}-7},
\label{bd.11}%
\end{equation}
we find that the Killing tensor that generates a quadratic in the momentum
conservation law of the form (\ref{bd.10}) is%
\begin{equation}
K_{ij}^{\left(  A\right)  }=a^{4}\phi\left(
\begin{tabular}
[c]{cc}%
$\phi$ & $\left(  1+\omega_{BD}\right)  a$\\
$\left(  1+\omega_{BD}\right)  a$ & $\left(  1+\omega_{BD}\right)  \frac
{a^{2}}{\phi}$%
\end{tabular}
\ \ \ \right)  , \label{bd.12}%
\end{equation}
with boundary%
\begin{equation}
\sigma_{A}\left(  a,\phi\right)  =-V_{2}\left(  \omega_{BD}+\frac{4}%
{3}\right)  a^{6}\phi^{-6\left(  1+\omega_{BD}\right)  }. \label{bd.13}%
\end{equation}

We proceed with the determination of the normal coordinates in which the field
equations can be solved by separation of variables. Performing the coordinate
transformation%
\begin{equation}
a=r^{\frac{2\left(  1+\omega_{BD}\right)  }{4+3\omega_{BD}}}\left(
\cosh\theta-\sinh\theta\right)  ^{\frac{2\left(  2\omega_{BD}+3\right)
}{\sqrt{3\left(  2\omega_{BD}+3\right)  }\left(  4+3\omega_{BD}\right)  }}~,
\label{bd.14}%
\end{equation}%
\begin{subequations}
\begin{equation}
\phi=r^{\frac{2}{4+3\omega_{BD}}}\left(  \cosh\theta-\sinh\theta\right)
^{\frac{6\left(  2\omega_{BD}+3\right)  }{\sqrt{3\left(  2\omega
_{BD}+3\right)  }\left(  4+3\omega_{BD}\right)  },} \label{bd.15}%
\end{equation}
with $\omega_{BD}\neq-\frac{3}{2},-\frac{4}{3}$, the Lagrangian of the field
equations (\ref{bd.07}) takes the following simple form,%
\end{subequations}
\begin{equation}
L\left(  r,\dot{r},\theta,\dot{\theta}\right)  =\frac{2\left(  2\omega
_{BD}+3\right)  }{\left(  4+3\omega_{BD}\right)  }\left(  \dot{r}^{2}%
-r^{2}\dot{\theta}^{2}\right)  +V_{eff}\left(  r,\theta\right)  ,
\label{bd.16}%
\end{equation}
where the effective potential is given by%
\begin{equation}
V_{eff}\left(  r,\theta\right)  =-V_{1}r^{2}+V_{2}\frac{\exp\left(
\frac{12\sqrt{\left(  2\omega_{BD}+3\right)  }}{\sqrt{3}}\theta\right)
}{r^{2}}. \label{bd.17}%
\end{equation}

The Lagrangian (\ref{bd.16}) describes the well-known Ermakov-Pinney dynamical
system in the $M^{2}$ spacetime. This means that the quadratic in the momentum
conservation law that follows from (\ref{bd.12}), (\ref{bd.13}) is the
Ermakov-Lewis invariant \cite{Aermakov}. Another way to observe this is to
study the point symmetries of the Lagrangian (\ref{bd.16}), which form an
$sl\left(  2,R\right)  $ Lie algebra.

From (\ref{bd.17}), we can write the first Friedmann's equations as follows
\begin{equation}
\frac{1}{2c_{0}}\left(  p_{r}^{2}-\frac{p_{\theta}^{2}}{r^{2}}\right)
-V_{1}r^{2}+V_{2}\frac{\exp\left(  \frac{12\sqrt{\left(  2\omega
_{BD}+3\right)  }}{\sqrt{3}}\theta\right)  }{r^{2}}=\mathcal{E\ }%
\label{bd.17a}%
\end{equation}
where $c_{0}=4\left(  2\omega_{BD}+3\right)  /\left(  4+3\omega_{BD}\right)  $
and%
\begin{equation}
\dot{r}=\frac{1}{c_{0}}p_{r}~,~r^{2}\dot{\theta}^{2}=\frac{1}{c_{0}}p_{\theta
},
\end{equation}
where the Ermakov-Lewis invariant is now%
\begin{equation}
\frac{1}{2c_{0}}p_{\theta}^{2}+V_{2}\exp\left(  \frac{12\sqrt{\left(
2\omega_{BD}+3\right)  }}{\sqrt{3}}\theta\right)  =I.
\end{equation}

We continue with the special case in which $\omega_{BD}=-\frac{4}{3}$. From
(\ref{bd.11}), now we have a linear potential $V\left(  \phi\right)  =\left(
V_{1}+V_{2}\right)  \phi$. \ The canonical coordinates, in which the system
can be solved with method of quadratures, are given by the transformation%
\begin{equation}
a=\left(  x+y\right)  ^{\frac{2}{3}}\exp\left(  y-x\right)  ~,~\phi=\left(
x+y\right)  ^{-1}\exp\left(  -3\left(  y-x\right)  \right)  ,\label{bd.18}%
\end{equation}
for which the Lagrangian becomes
\begin{equation}
L\left(  x,\dot{x},y,\dot{y}\right)  =-\dot{x}^{2}+\dot{y}^{2}+\left(
V_{1}+V_{2}\right)  \left(  x+y\right)  .\label{bd.19}%
\end{equation}
In contrast to the case above, the linear potential is that of a constant
force, $V_{1}r^{2},$ rather than an oscillator. \ 

Let $V_{12}=-\left(  V_{1}+V_{2}\right)  $, then we have the analytical
solution
\begin{equation}
x=\frac{V_{12}}{2}t^{2}+x_{1}t+x_{0}~,~y\left(  t\right)  =-\frac{V_{12}}%
{2}t^{2}+y_{1}t+y_{0}, \label{bd.20}%
\end{equation}
or, for the scale factor:%
\begin{equation}
a\left(  t\right)  =\left(  \left(  x_{1}+y_{1}\right)  t+\left(  x_{0}%
+y_{0}\right)  \right)  ^{\frac{2}{3}}\exp\left(  -V_{12}t^{2}+\left(
y_{1}-x_{1}\right)  t\right)  e^{y_{0}-x_{0}}, \label{bd.21}%
\end{equation}
with $a\left(  t\rightarrow0\right)  \equiv\left(  x_{0}+y_{0}\right)
^{\frac{2}{3}}e^{y_{0}-x_{0}}$. For large $t,$ with $V_{12}<0$, we have%
\begin{equation}
a\left(  t\right)  \simeq t^{\frac{2}{3}}\exp\left(  \left\vert V_{12}%
\right\vert t^{2}\right)  . \label{bd.22}%
\end{equation}

Eqn. (\ref{bd.22}) provides that the effective fluid is described by the
energy momentum tensor $T_{\mu\nu}=\left(  \mu_{eff}+p_{eff}\right)  u_{\mu
}u_{\nu}+p_{eff}g_{\mu\nu}$, where%
\begin{equation}
\mu_{eff}\simeq\frac{1}{3t^{2}}+2\left\vert V_{12}\right\vert +3\left(
\left\vert V_{12}\right\vert \right)  ^{2}t^{2}%
\end{equation}%
\begin{equation}
p_{eff}\simeq\frac{1}{3t^{2}}-4\left\vert V_{12}\right\vert -3\left(
\left\vert V_{12}\right\vert \right)  ^{2}t^{2}%
\end{equation}
where for large values of $t$ the above set of equations reduce to%
\begin{equation}
\mu_{eff}\simeq2\left\vert V_{12}\right\vert +3\left(  \left\vert
V_{12}\right\vert \right)  ^{2}t^{2}~
\end{equation}
and%
\begin{equation}
p_{eff}\simeq-\mu_{eff}-2\left\vert V_{12}\right\vert .
\end{equation}
where the latter equation resembles that of a cosmological constant, i.e.,
\ $p_{eff}\simeq-(\mu_{eff}+2\left\vert V_{12}\right\vert )=-\rho_{eff}$.

\begin{figure}[t]
\centering\includegraphics[scale=0.45]{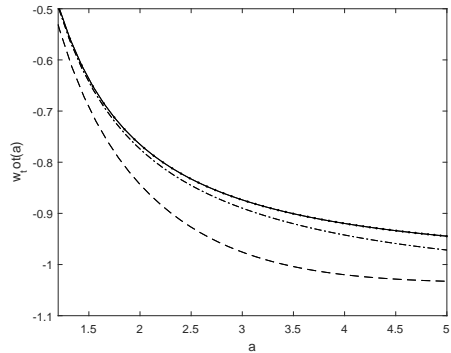}
\caption{The behavior of $w_{eff}\left(  a\right)  $ as a function of the
scale factor (\ref{bd.22}). We can see that $w_{eff}<-1$, while for large
values of $a$,~$w_{eff}$ increases such that $w_{eff}\rightarrow-1$. The solid
line is for $\left\vert V_{12}\right\vert =10^{-4}$, dash-dot line is for
$\left\vert V_{12}\right\vert =10^{-3}$, dash-dash line is for $\left\vert
V_{12}\right\vert =10^{-2}$, and dot-dot line is for $\left\vert
V_{12}\right\vert =10^{-1}.$}%
\label{fig011}%
\end{figure}

Furthermore, for the deceleration parameter we find that
\begin{equation}
q=-1+\frac{3}{2}\left(  1-3\left\vert V_{12}\right\vert t^{2}\right)  \left(
1+3\left\vert V_{12}\right\vert t^{2}\right)  ^{-2}%
\end{equation}
Obviously, we observe that although at late enough times we recover the de
Sitter solution $(q=-1)$ the total equation of state parameter can cross the
phantom line (see fig. \ref{fig011} ) because the scalar curvature diverges to
the future, with $R\rightarrow\frac{\ddot{a}}{a}\rightarrow O(t^{2}%
)$.

\subsection{ Potential B}

For a potential of the form%
\begin{equation}
V_{A}\left(  \phi\right)  =V_{1}\phi+V_{2}\phi^{P}, \label{bd.23}%
\end{equation}
where
\begin{equation}
P=\frac{3}{2}-\frac{\sqrt{3\left(  2\omega_{BD}+3\right)  }}{2}, \label{bd.24}%
\end{equation}
the Killing tensor which generates quadratic in the momentum conservation law
is%
\begin{equation}
K_{ij}^{\left(  B\right)  }=a^{q_{1}}\phi^{q_{2}}\left(
\begin{tabular}
[c]{cc}%
$3\frac{\left(  q_{3}+\sqrt{3q_{3}}\right)  }{\omega_{BD}}^{2}\phi$ &
$-\frac{\sqrt{3q_{3}}\left(  q_{3}+\sqrt{3q_{3}}\right)  }{\omega_{BD}}$\\
$-\frac{\sqrt{3q_{3}}\left(  q_{3}+\sqrt{3q_{3}}\right)  }{\omega_{BD}}$ &
$\frac{a^{2}}{\phi}q_{3}$%
\end{tabular}
\ \ \ \ \ \right)  , \label{bd.25}%
\end{equation}
and the boundary term is%
\begin{equation}
\sigma_{B}\left(  a,\phi\right)  =a^{2+q_{1}}\phi^{1+q_{2}}\left(
\frac{\left(  q_{3}+\sqrt{3q_{3}}\left(  1+\omega_{BD}\right)  \right)
}{\omega_{BD}\left(  q_{2}-1\right)  }\right)  , \label{bd.26}%
\end{equation}
where the constants~$q_{1},q_{2}$, and $q_{3}$ are~$q_{1}=1+\sqrt{\frac
{3}{q_{3}}}~,~q_{2}=\sqrt{\frac{3}{q_{3}}}\left(  1+\omega_{BD}\right)  ,~$and
$q_{3}=2\omega_{BD}+3$.

The normal coordinates are defined by the relations%
\begin{equation}
a=\left(  uv\right)  ^{\frac{1+q_{2}}{\sqrt{\frac{3}{q_{3}}}\left(
4+3\omega_{BD}\right)  }}\left(  v\right)  ^{\frac{2q_{3}}{\sqrt{3q_{3}%
}\left(  4+3\omega_{BD}\right)  },} \label{bd.27}%
\end{equation}%
\begin{equation}
\phi=\left(  uv\right)  ^{\frac{1+\sqrt{3q_{3}}}{\left(  4+3\omega
_{BD}\right)  }}\left(  v\right)  ^{\frac{6\sqrt{q_{3}}}{\sqrt{q_{3}}\left(
4+3\omega_{BD}\right)  }}, \label{bd.28}%
\end{equation}
and the Lagrangian becomes%
\begin{equation}
L\left(  u,\dot{u},v,\dot{v}\right)  =-\frac{2q_{3}}{\left(  4+3\omega
_{BD}\right)  }\dot{u}\dot{v}-V_{1}uv+V_{2}\left(  v\right)  ^{\frac
{\sqrt{3q_{3}}-3q_{3}}{4+3\omega_{BD}}}. \label{bd.29}%
\end{equation}
Hence, we find that the field equations are%
\begin{equation}
\ddot{u}-\bar{V}_{1}u+\bar{V}_{2}v^{k}=0~,~\ddot{v}-\bar{V}_{1}v=0,
\label{bd.30}%
\end{equation}
where $\bar{V}_{1}=\bar{V}_{1}\left(  V_{1},\omega_{BD}\right)  $ and $\bar
{V}_{2}=\bar{V}_{2}\left(  V_{2},\omega_{BD}\right)  $ and $k=\frac
{\sqrt{3q_{3}}-3q_{3}}{4+3\omega_{BD}}-1$.

From (\ref{bd.30}) we have that~%
\begin{equation}
v\left(  t\right)  =v_{1}e^{\sqrt{\bar{V}_{1}}t}+v_{2}e^{-\sqrt{\bar{V}_{1}}t}%
\end{equation}
while the $u$ parameter   satisfies the linear equation%
\begin{equation}
\ddot{u}-\bar{V}_{1}u+\left(  v_{1}e^{\sqrt{\bar{V}_{1}}t}+v_{2}e^{-\sqrt
{\bar{V}_{1}}t}\right)  ^{k}=0
\end{equation}
a solution of which is%
\begin{equation}
u\left(  t\right)  =u_{1}e^{\sqrt{\bar{V}_{1}}t}+u_{2}e^{-\sqrt{\bar{V}_{1}}%
t}+\frac{e^{-\sqrt{V_{1}t}}}{2\sqrt{V_{1}}}\int e^{\sqrt{V_{1}t}}\left(
v\left(  t\right)  \right)  ^{k}dt-\frac{e^{\sqrt{V_{1}t}}}{2\sqrt{V_{1}}}\int
e^{-\sqrt{V_{1}t}}\left(  v\left(  t\right)  \right)  ^{k}dt\label{bd.32}%
\end{equation}
Notice, that equation (\ref{bd.001}) provides an additional algebraic relation
for the integration constants. \ 

For $\bar{V}_{1}>0$ and large values of $t$, we have that $v\left(  t\right)
\simeq v_{1}e^{\sqrt{\bar{V}_{1}}t}$ and consequently from (\ref{bd.32}) it
follows that $u\left(  t\right)  $ behaves like%
\begin{equation}
u\left(  t\right)  =u_{1}e^{\sqrt{\bar{V}_{1}}t}+\mu e^{\Omega t}%
\end{equation}
where $\mu=\left(  V_{1},k,v_{1}\right)  $ and $\Omega=\Omega\left(
k,V_{1}\right)  $. Hence either for large values of $t$, the $u\left(
t\right)  $ will be exponential as also from (\ref{bd.27}) the scale factor
which means that the de Sitter universe is approached.

In the special case of $\omega_{BD}=-\frac{4}{3}$, the normal coordinates are
given by the transformation%
\begin{equation}
a=u^{\frac{2}{3}}e^{-v}~,~\phi=u^{-1}e^{3v}.
\end{equation}
Therefore, the Lagrangian of the field equations takes the simple form
\begin{equation}
L\left(  u,\dot{u},v,\dot{v}\right)  =-2\dot{u}\dot{v}+V_{1}u+V_{2}e^{3v}.
\end{equation}

In this context the Euler-Lagrange equations are
\begin{equation}
\ddot{u}+\frac{3}{2}V_{2}e^{3v}=0~,~\ddot{v}+\frac{V_{1}}{2}=0,
\end{equation}
from which we find $v\left(  t\right)  =\frac{V_{1}}{4}t^{2}+v_{1}t+v_{0}$ and%
\begin{equation}
\dot{u}=-\frac{3}{2}V_{2}e^{v_{0}}\int\exp\left(  \frac{V_{1}}{4}t^{2}%
+v_{1}t\right)  dt.
\end{equation}

\subsection{Potentials C$~$and D~}

Furthermore, a quadratic in the momentum conservation law generated by the
Killing tensors $K_{ij}=K_{(J}\otimes K_{3)}$exists for the field equations
when the potential is \
\begin{equation}
V\left(  \phi\right)  =V_{1}\phi^{p_{+}^{J}}+V_{2}\phi^{p_{-}^{J}},
\end{equation}
with
\begin{equation}
p_{\pm}^{1}=\frac{13-5\sqrt{3q_{3}}\pm\sqrt{3q_{3}+1-2\sqrt{3q_{3}}}}%
{8},~\text{for~}J=1,
\end{equation}
and%
\begin{equation}
p_{\pm}^{2}=\frac{13+5\sqrt{3q_{3}}\pm\sqrt{3q_{3}-4+\sqrt{3q_{3}}}}%
{8}~,~\text{for }J=2,
\end{equation}
where $K_{J}$ are the two gradient isometries of the minisuperspace
(\ref{bd.08}), and $K_{3}$ is the non-gradient isometry.

The normal coordinates are given by the transformation (\ref{bd.27}),
(\ref{bd.28}). In the normal coordinates the effective potential is $V\left(
u,v\right)  =u^{\bar{P}_{1}}v^{\bar{P}_{2}}$. Using the Hamilton-Jacobi
equation and the second conservation law, the field equations can be reduced
to a system of nonlinear autonomous first-order differential equations. The
dynamical system is one of the Lie integrable systems which have been
classified in \cite{dask}. The results can be compared with the power-law
$V(\phi)$ solutions found in ref. \cite{bmaeda}.

The analysis that we have presented here holds for $\omega_{BD}\neq0$. In the
limit $\omega_{BD}=0$, we can see that only the potential A provides
integrability for the $f\left(  R\right)  $-gravity \cite{kt2}. We continue
our analysis with the study of the evolution of the field equations for the
integrable potentials. \emph{ }

\section{Dynamical evolution}

\label{section4}

We continue with the study of the critical points of the gravitational field
equations. We will perform that analysis for a general potential, $V\left(
\phi\right)  $, so that we can determine the fixed points for the integrable
potentials found in the previous sections.

First, we introduce the dimensionless variables \ (for instance see
\cite{ref1,ref2,ref3,ref4})
\begin{equation}
x=\frac{\dot{\phi}}{H\phi}~,~y=\sqrt{\frac{V}{3\phi}}\frac{1}{H}. \label{xvar}%
\end{equation}
and the new lapse function $N=\ln\left(  a\right)  $.

In the vacuum scenario, using the new variables, the gravitational field
equations (\ref{bd.001})-(\ref{bd.003}) become%
\begin{equation}
0=1+x-\frac{\omega_{BD}}{6}x^{2}-y^{2}, \label{om1}%
\end{equation}
and%
\begin{equation}
\frac{dx}{dN}=-3x-x^{2}-x\frac{\dot{H}}{H^{2}}+3(1+x-\frac{\omega_{BD}}%
{6}x^{2}-y^{2})\frac{2+\omega_{BD}}{3+2\omega_{BD}}+\frac{6}{3+2\omega_{BD}%
}y^{2}(2+\lambda), \label{x1}%
\end{equation}%
\begin{equation}
\frac{dy}{dN}=-y[\frac{\dot{H}}{H^{2}}+\frac{1}{2}x(1+\lambda)], \label{y1}%
\end{equation}%
\begin{equation}
\frac{d\lambda}{dN}=x\lambda\lbrack1-\lambda(\Gamma\left(  \lambda\right)
-1)], \label{l1}%
\end{equation}
where
\begin{equation}
\lambda=-\phi\frac{V_{,\phi}}{V}~,~\Gamma\left(  \lambda\right)
=\frac{V_{,\phi\phi}V}{\left(  V_{,\phi}\right)  ^{2}}. \label{lvar}%
\end{equation}
\ ~The dynamical system (\ref{om1})-(\ref{l1}) is a system of first-order
differential equations. Also, from (\ref{om1}) we find the constraint%
\begin{equation}
y^{2}=1+x-\frac{\omega_{BD}}{6}x^{2}. \label{omegam}%
\end{equation}

Hence, the dynamical system (\ref{x1})-(\ref{l1}) can be reduced to\ the
following system
\begin{equation}
x^{\prime}=-x[3(1+x)-\frac{\omega_{BD}}{2}x^{2}-\frac{3}{3+2\omega_{BD}%
}(1+x-\frac{\omega_{BD}}{6}x^{2})(2+\lambda)], \label{dec2}%
\end{equation}
and%
\begin{equation}
\lambda^{\prime}=-x(\bar{\Gamma}(\lambda)), \label{dec3}%
\end{equation}
where%

\begin{equation}
\bar{\Gamma}(\lambda)=\lambda\lbrack1-\lambda(\Gamma\left(  \lambda\right)
-1)].
\end{equation}

However, from (\ref{lvar}), we see that for $V\left(  \phi\right)  =V_{0}%
\phi^{A}$, $\lambda=-A$ , $\Gamma=(A-1)/A,~\bar{\Gamma}=0$ and so (\ref{dec3})
is identically constant and only equation (\ref{dec2}) survives.

Moreover, it is important to mention that all the cosmological parameters can
be expressed in terms of the new variables. As an example, the deceleration
parameter $q$ is given by%
\begin{align}
q  &  =-1+3(1+x-\frac{\omega_{BD}}{6}x^{2}-y^{2})\frac{\omega_{BD}+2}%
{(2\omega_{BD}+3)}+\nonumber\\
&  ~~~~~~~~\ ~~~\ ~~~~~~~~~~~~~~~~~~~~~~~~+\frac{\omega_{BD}}{2}x^{2}%
-2x+\frac{3(\lambda+2)y^{2}}{2\omega_{BD}+3}. \label{dec1}%
\end{align}

\section{Power-law potential}

We continue our analysis by considering that the potential is power-law. We
study separate the case in which the potential is quadratic.

\subsubsection{Quadratic potential}

As we have already shown above for $V\left(  \phi\right)  =V_{0}\phi^{2}$, we
have $\lambda=-2$, which implies that equation (\ref{dec2}) is simplified to
\begin{equation}
x^{\prime}=-3x[(1+x)-\frac{\omega_{BD}}{6}x^{2}], \label{dec2V2}%
\end{equation}
where the corresponding critical points are
\begin{equation}
P_{1}:x=0~;~P_{2}^{\pm}:x=\frac{3\pm\sqrt{3\left(  2\omega_{BD}+3\right)  }%
}{\omega_{BD}}. \label{ss101}%
\end{equation}

At the point $P_{1}$ we find that the deceleration parameter is $q\left(
P_{1}\right)  =-1$, which means that $w_{\phi}=-1$, hence the Brans-Dicke
field is acting like a cosmological constant. On the other hand points
$P_{2}^{\pm}$ are real when $\omega_{BD}>-\frac{3}{2}$, and the deceleration
parameter then takes the following forms:
\begin{equation}
q(P_{2}^{+})=\frac{\left(  2\omega_{BD}+3\right)  +\sqrt{3\left(  2\omega
_{BD}+3\right)  }}{\omega_{BD}}, \label{ss1}%
\end{equation}%
\begin{equation}
q(P_{2}^{-})=\frac{\left(  2\omega_{BD}+3\right)  -\sqrt{3\left(  2\omega
_{BD}+3\right)  }}{\omega_{BD}}. \label{ss2}%
\end{equation}
One can easily check that $P_{2}^{+}$ describes an accelerating universe,
since $q\left(  P_{2}^{+}\right)  <0$, for $\omega_{BD}\in\left(  -\frac{3}%
{2},0\right)  .$

As far as the stability of these points is concerned, this can be checked
easily by studying the derivative of the right-hand part, $\mathbf{F}\left(
x\right)  ,$ of (\ref{dec2V2}). Thus, we find
\begin{equation}
\frac{d\mathbf{F}\left(  x\right)  }{dx}|_{P_{1}}=-3,
\end{equation}%
\begin{equation}
\frac{d\mathbf{F}\left(  x\right)  }{dx}|_{P_{2}^{+}}=\frac{3}{\omega_{BD}%
}\left(  \left(  2\omega_{BD}+3\right)  +\sqrt{3\left(  2\omega_{BD}+3\right)
}\right)  ,
\end{equation}%
\begin{equation}
\frac{d\mathbf{F}\left(  x\right)  }{dx}|_{P_{2}^{-}}=\frac{3}{\omega_{BD}%
}\left(  \left(  2\omega_{BD}+3\right)  -\sqrt{3\left(  2\omega_{BD}+3\right)
}\right)  .
\end{equation}

Therefore, point $P_{1}$ is always a stable attractor, while point $P_{2}^{+}$
is stable when $\omega_{BD}\in\left(  -\frac{3}{2},0\right)  $ and point
$P_{2}^{-}$ is always unstable.

From the potentials of the previous section, and for $\omega_{BD}\neq-\frac
{3}{2},0,$ we see that potential B, eqn. (\ref{bd.23}), becomes a quadratic
potential for $\omega_{BD}=-\frac{4}{3}$ and $V_{1}=0$, which means that it
admits two accelerated stable points: $P_{1}$ and $P_{2}^{+}$. \ For the
potentials C and D, we see that only potential C can describe a quadratic
potential for $V_{2}=0$ and $\omega_{BD}=-\frac{4}{3}$, and for $V_{1}=0$ and
$\omega_{BD}=-\frac{40}{27}$. This implies that in both cases two stable
points exist and they each describe an accelerating universe.

\subsubsection{Potential $V\left(  \phi\right)  =V_{0}\phi^{A}$}

For a general power-law potential $V\left(  \phi\right)  =V_{0}\phi^{A}$, with
$A\neq-2$, we find that $\lambda=-A$ and $\bar{\Gamma}=0$. Hence, for the
differential equation (\ref{dec2}) we find the critical points $P_{1}~$and
$P_{2}^{\pm}$ of (\ref{ss101}). We would like to stress here that although the
points between (\ref{dec2}) and (\ref{ss101}) are the same the corresponding
physical parameters, such that the deceleration parameter and the stability of
the critical points are different because they depend on the value of the
power $A$.

As far as the deceleration parameters are concerned we find that
\begin{equation}
q\left(  P_{1}\right)  =\frac{6\left(  1-A\right)  -4\omega_{BD}}{2\left(
2\omega_{BD}+3\right)  },
\end{equation}
and%
\begin{equation}
q\left(  P_{2}^{\pm}\right)  =\frac{\left(  2-A\right)  \pm\sqrt{\left(
2\omega_{BD}+3\right)  }}{\omega_{BD}}.
\end{equation}

Hence $P_{1}$ is an accelerated point as long as%
\begin{equation}
\omega_{BD}<-\frac{3}{2}~,~A\leq\frac{3-2\omega_{BD}}{3}~\text{or~}\omega
_{BD}>-\frac{3}{2}~,~A>\frac{3-2\omega_{BD}}{3},
\end{equation}
while points $P_{2}^{\pm}$ provide an accelerated universe when
\begin{equation}
\omega_{BD}\in\left(  -\frac{3}{2},0\right)  ~,~A\leq2\pm\sqrt{3+2\omega_{BD}%
}~~\text{or~}\omega_{BD}>0~\,,~A>2\pm\sqrt{3+2\omega_{BD}}.
\end{equation}

It is easy to see that $P_{1}$ describes a de Sitter point ($q=-1$) only when
$A=2$ that is, the quadratic potential. On the other hand, points $P_{2}^{\pm
}$ can describe de Sitter phases for $A=2+\omega_{BD}\pm\sqrt{3+2\omega_{BD}}%
$. Concerning the stability of the aforementioned points the situation is the
following:$P_{1}$ is always stable as long as
\begin{equation}
\omega_{BD}<-\frac{3}{2}~~,~A<-\left(  1+2\omega_{BD}\right)  \text{ or
}\omega_{BD}>\frac{3}{2}~,~A>-\left(  1+2\omega_{BD}\right)  ,
\end{equation}
while $P_{2}^{\pm}$ are stable when
\begin{equation}
\omega_{BD}\in\left(  -\frac{3}{2},0\right)  ~,~A<-\left(  1+2\omega
_{BD}\right)  \text{ or }\omega_{BD}>0\text{~,~}A>-\left(  1+2\omega
_{BD}\right)  .
\end{equation}

\subsection{General potential}

Consider now a general potential, $V\left(  \phi\right)  $, which means that
we have a general function $\bar{\Gamma}\left(  \lambda\right)  $. If we
assume that $\lambda=\lambda_{1}$ is a solution of the algebraic equation
$\bar{\Gamma}\left(  \lambda\right)  =0$, then we find that the system
(\ref{dec2})-(\ref{dec3}) admits the following critical points
\begin{equation}
P_{1}:x=0~,~\lambda=-A,
\end{equation}%
\begin{equation}
P_{2}^{\pm}\left(  \lambda_{1}\right)  :~x=\frac{3\pm\sqrt{3\left(
2\omega_{BD}+3\right)  }}{\omega_{BD}}~,~\lambda=\lambda_{1},
\end{equation}
and
\begin{equation}
P_{3}\left(  \lambda_{1}\right)  :x=\frac{2(\lambda_{1}+2)}{1+2\omega
_{BD}-\lambda_{1}}~,~\lambda=\lambda_{1}.
\end{equation}
From the latter, we observe that for $\lambda_{1}=-2$, $P_{3}$ reduces to
$P_{1}$. On the other hand, when $\lambda_{1}=2\omega_{BD}+1$, point $P_{3}$
does not exist and only points $P_{2}^{\pm}$ exist.

The linearization of the system (\ref{dec2})-(\ref{dec3}) around the point
$P_{1}$ provides the following eigenvalues
\begin{equation}
e^{\pm}\left(  P_{1}\right)  =-\frac{3}{2}\pm\sqrt{3\left(  2\omega
_{BD}+3\right)  [16\bar{\Gamma}\left(  -A\right)  +3\left(  2\omega
_{BD}+3\right)  ]}.
\end{equation}
When $\left(  2\omega_{BD}+3\right)  <0$, or $\left(  16\bar{\Gamma}\left(
-2\right)  +3\left(  2\omega_{BD}+3\right)  \right)  <0$, the eigenvalues have
negative real part so $P_{1}$ describes a stable spiral. On the other hand, if
the condition $e^{+}\left(  P_{1}\right)  e^{-}\left(  P_{1}\right)  >0$ is
satisfied, then%
\begin{equation}
\frac{12\bar{\Gamma}\left(  -A\right)  }{2\omega_{BD}+3}<0,
\end{equation}
and $P_{1}$ is stable. Lastly, point $P_{1}$ describes a de Sitter point,
because $w_{tot}=-1.$

Concerning the eigenvalues of point $P_{3}$ we have%
\begin{equation}
e_{1}\left(  P_{3}\right)  =-\frac{6\omega_{BD}+5-\lambda_{1}\left(
\lambda_{1}+4\right)  }{1+2\omega_{BD}-\lambda_{1}}~,~e_{2}\left(
P_{3}\right)  =-\frac{2\lambda_{1}\left(  2+\lambda_{1}\right)  \bar{\Gamma
}_{,\lambda}\left(  \lambda_{1}\right)  }{1+2\omega_{BD}-\lambda_{1}}.
\end{equation}
If $\bar{\Gamma}_{,\lambda}\left(  \lambda_{1}\right)  >0$ \ then we find that
$P_{3}$ is stable when%
\begin{equation}
\lambda_{1}>0,~\omega_{BD}>-\frac{1}{2}\text{ and }1+2\omega_{BD}-\lambda
_{1}>0,
\end{equation}
or%
\begin{equation}
\lambda_{1}<-2~,~\text{and }1+2\omega_{BD}-\lambda_{1}>0,
\end{equation}
or%
\begin{equation}
-2<\lambda_{1}<0,~\omega_{BD}<-\frac{1}{2}\text{ and }1+2\omega_{BD}%
-\lambda_{1}<0.
\end{equation}
Alternatively, for $\bar{\Gamma}_{,\lambda}\left(  \lambda_{1}\right)  <0$,
$P_{3}$ is stable when%
\begin{equation}
\lambda_{1}<-2~,~\omega_{BD}<-\frac{3}{2}~,~\text{and }1+2\omega_{BD}%
-\lambda_{1}<0,
\end{equation}
or%
\begin{equation}
\lambda_{1}>0~\text{and}~1+2\omega_{BD}-\lambda_{1}<0,
\end{equation}
or%
\begin{equation}
-2<\lambda_{1}<0~,~\omega_{BD}>-\frac{3}{2}~,~\text{and }1+2\omega
_{BD}-\lambda_{1}>0.
\end{equation}

Finally, for points $P_{2}^{\pm}$, we compute the corresponding eigenvalues%
\begin{equation}
~e_{1}\left(  P_{2}^{\pm}\right)  =\frac{3\left(  2\omega_{BD}+3\right)
\pm\sqrt{3\left(  2\omega_{BD}+3\right)  }-\left(  3\pm\sqrt{3\left(
2\omega_{BD}+3\right)  }\right)  \lambda_{1}}{\omega_{BD}},
\end{equation}
and%
\begin{equation}
~e_{2}\left(  P_{2}^{\pm}\right)  =-\frac{3\pm\sqrt{3\left(  2\omega
_{BD}+3\right)  }\lambda_{1}\bar{\Gamma}_{,\lambda}\left(  \lambda_{1}\right)
}{\omega_{BD}}.
\end{equation}
Recall that points $P_{2}^{\pm}$ exist when $\omega_{BD}>-\frac{3}{2}$ and
$\omega_{BD}\neq0$. \ At these points the deceleration parameters are
calculated to be (\ref{ss1}) and (\ref{ss2}) respectively. Hence, for
$\omega_{BD}\in\left(  -\frac{3}{2},0\right)  $,~the point $P_{2}^{+}$
describes an accelerating universe.

\subsection{Specific potentials}

Consider now the potential~%
\begin{equation}
V_{1}\left(  \phi\right)  =V_{1}\phi+V_{2}\phi^{2}. \label{pot02}%
\end{equation}

$~$From eq.(\ref{lvar}) we find $\phi=-\frac{V_{1}}{V_{2}}\left(
1+\lambda\right)  \left(  2+\lambda\right)  ^{-1}$, where $\bar{\Gamma}\left(
\lambda\right)  =\left(  \lambda+1\right)  \left(  \lambda+2\right)  $, with
solutions $\lambda=-1$, and $\lambda=-2$. \ Therefore, in this case there are
six critical points. Points $P_{1}$,~$P_{3}$ for $\lambda_{1}=-1$, and the
four points $P_{2}^{\pm}\left(  -1\right)  $,~$P_{2}^{\pm}\left(  -2\right)  $.

The point $P_{3}$ is always stable for $\omega_{BD}>-1$, and the deceleration
parameter is $q\left(  P_{3}\right)  =-\frac{\left(  2+\omega_{BD}\right)
\left(  2\omega_{BD}+3\right)  }{2\left(  1+\omega_{BD}\right)  ^{2}}$, which
gives that $q\left(  P_{3}\right)  <0$, when $\omega_{BD}\in\left(
-\infty,-2\right)  \cup\left(  -\frac{3}{2},-1\right)  \cup\left(
-1,+\infty\right)  $. Therefore, in the special case of $\omega_{BD}=-\frac
{4}{3}$, we have that $q\left(  P_{3}\right)  =-1$. \ Among the points
$P_{2}^{\pm},$ only the $P_{2}^{+}\left(  -2\right)  $ and $P_{2}^{+}\left(
-1\right)  ~$can be stable \ for $\omega_{BD}\in\left(  -\frac{3}{2},0\right)
$, and $\omega_{BD}\in\left(  -\frac{4}{3},0\right)  $ respectively. The other
two points are unstable, while the deceleration parameters are given by eqns.
(\ref{ss1}), (\ref{ss2}).

Despite the fact that eq.(\ref{pot02}) is a simple generalization of the
quadratic potential, we can see that new critical points appear in this
dynamical system. Furthermore, for the special value $\omega_{BD}=-\frac{4}%
{3}$ , $P_{3}$ is stable and it describes a de Sitter universe. This is the
value at which the potential B, eqn.$~$(\ref{bd.23}), takes the form of
(\ref{pot02}). Hence, as in the case of a minimally coupled scalar field
\cite{anb}, symmetries provide us with models which can describe the de Sitter
phase of the universe.

For a second application we consider the potential%
\begin{equation}
V_{2}\left(  \phi\right)  =V_{1}\phi+V_{2}\phi^{-1},\text{ } \label{pot.02}%
\end{equation}
from where we calculate that $\bar{\Gamma}\left(  \lambda\right)
=(\lambda^{2}-1),$ with solutions $\lambda=\pm1.$Therefore, we find seven
critical points, $P_{1},~P_{3}\left(  -1\right)  ,~P_{3}\left(  1\right)
\,,~P_{2}^{\pm}\left(  -1\right)  \,~$and $P_{2}^{\pm}\left(  1\right)  $.

Point $P_{1}$ is stable for $\omega_{BD}>-\frac{3}{2}$, and it describes a de
Sitter universe. Point $P_{3}\left(  -1\right)  $ is stable for $\omega
_{BD}>-1$ while the stability of $P_{3}\left(  1\right)  $ holds for
$\omega_{BD}>0$. At $P_{3}\left(  -1\right)  $, the deceleration parameters is
$q\left(  P_{3}\left(  -1\right)  \right)  =-\frac{\left(  2+\omega
_{BD}\right)  \left(  2\omega_{BD}+3\right)  }{2\left(  1+\omega_{BD}\right)
^{2}}~$as above, while $q\left(  P_{3}\left(  1\right)  \right)  =-1-\frac
{3}{2\omega_{BD}}$, which shows that $q\left(  P_{3}\left(  1\right)  \right)
<0$, when $\omega_{BD}\in\left(  -\infty,-\frac{3}{2}\right)  \cup\left(
0,+\infty\right)  $. Therefore, when the points $P_{3}$ are stable they
provide an accelerated cosmological expansion. Moreover, $P_{2}^{+}\left(
-1\right)  $ is stable when $\omega_{BD}\in\left(  -\frac{4}{3},0\right)  $
while the rest of the points are always unstable. \ It is interesting to
mention here that potential A with $\omega_{BD}=1$ and potential B with
$\omega_{BD}=\frac{8}{3}$, reduce to potential (\ref{pot.02}).

\begin{figure}[t]
\centering\includegraphics[scale=0.59]{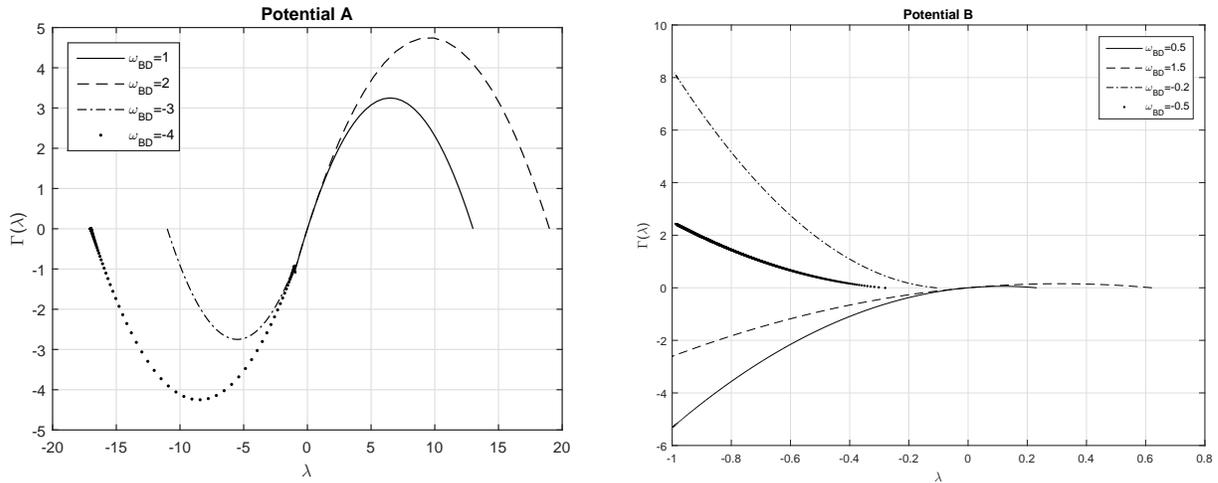}
\centering\includegraphics[scale=0.45]{potB.eps}
\caption{Qualitative evolution of $\lambda-\bar{\Gamma}\left(  \lambda\right)
$ for the potentials A and B with various values of the Brans-Dicke parameter
$\omega_{BD}$. For the plots we have assumed that the two constants
$V_{1},V_{2}~$ are equal and positive. The plots are for $\phi\in(0,10^{3})$.}%
\label{figpot}%
\end{figure}

In fig. \ref{figpot} we present the qualitative evolution of $\bar{\Gamma
}\left(  \lambda\right)  $ function for the potentials A~and B in the case
that the two constants $V_{1},V_{2}$ are equal and for positive values of the
scalar field $\phi$. We observe that for positive Brans-Dicke parameters there
are at least two solutions $\lambda_{1},~$for the algebraic equation
$\bar{\Gamma}\left(  \lambda_{1}\right)  =0.$

\section{Conclusions}

\label{conclusions}

The method of group invariant transformations and specifically the existence
of contact symmetries for the field equations in the Brans-Dicke cosmology is
the main focus of this work. We determined the unknown functional form of the
Brans-Dicke self-interaction potential $V(\phi)$ by assuming that a quadratic
conservation law exists.
In the case of a FLRW geometry the existence of a second conservation law
leads to integrable field equations, and thus, the cosmological solution can
be found by quadratures.

We determined four families of power-law potentials in which the powers depend
on the value of the Brans-Dicke parameter. From these potentials only one
survives for $\omega_{BD}=0$, and recover a specific theory of $f\left(
R\right)  $-gravity. Furthermore, we studied the critical points of the field
equations in the vacuum scenario, using the dimensionless variables. In these
variables the field equations reduce to a system of algebraic-differential
equations of first order. Utilizing the algebraic equations, the final system
contains two first-order ordinary differential equations, while in the limit
of quadratic potential the system reduces to one first-order differential equation.

For the quadratic form of the potential $V(\phi)$ we found three critical
points, of which one always describes the de Sitter universe and it is stable
for $\omega_{BD}>-\frac{3}{2}$. For the general potential $V\left(
\phi\right)  $, we find that the maximum number of critical points is $1+3n,$
where $n$ is the number of solutions of the algebraic equation $\bar{\Gamma
}\left(  \lambda\right)  =0$.

Finally, in order to demonstrate our results we considered two special forms
of the potential and studied the stability of the critical points and the
physical parameters. Two of the power-law families that we calculated from the
application of the Killing tensors reduce to those special potentials and we
derived the corresponding stable points which lead to an accelerated expansion
of the universe. This is an interesting result because it coincides with
previous results from other gravitational theories \cite{bshaw}.

\begin{acknowledgments}
JDB is supported by the STFC of the United Kingdom. AG acknowledges financial
support\ of FONDECYT grant no. 1150246. AP acknowledges financial support\ of
FONDECYT grant no. 3160121.
\end{acknowledgments}


\begin{thebibliography}{99}                                                                                               %


\bibitem {Teg}M. Tegmark et al., Astrophys. J. \textbf{606,} 702 (2004)

\bibitem {Kowal}M. Kowalski et al., Astrophys. J. \textbf{686,} 749 (2008)

\bibitem {Komatsu}E. Komatsu et al., Astrophys. J. Suppl. Ser. \textbf{180},
330 (2009)

\bibitem {suzuki11}N. Suzuki et. al., Astrophys. J. \textbf{746,} 85 (2012)

\bibitem {Ade15}Planck Collaboration: P.A.R. Ade et al., A\&A \textbf{594,}
A13 (2016)

\bibitem {Aref1}H.A. Buchdahl, Mon. Not. R. astron. Soc. \textbf{150,} 1 (1970)

\bibitem {Aref2}G.R. Bengochea and R. Ferraro, Phys. Rev. D \textbf{79,}
124019 (2009)

\bibitem {Aref3}V. Faraoni, Int. J. Mod. Phys. D \textbf{11,} 471 (2002).

\bibitem {Aref4}A.A. Starobinsky, Phys. Lett. B \textbf{91,} 99 (1980).

\bibitem {Aref5}R.~Ferraro and F.~Fiorini, Phys.\ Rev.\ D \textbf{75,} 084031 (2007)

\bibitem {Aref6}J.A.S. Lima, S. Basilakos and J. Sola, Mon. Not. R. astron.
Soc. \textbf{431}, 923 (2013).

\bibitem {Ar1}B. Ratra and P.J.E. Peebles, Phys. Rev D. \textbf{37,} 3406 (1988)

\bibitem {Ar2}J.M. Overduin and F.I. Cooperstock, Phys. Rev. D \textbf{58,}
043506 (1998)

\bibitem {Ar3}S. Basilakos, M. Plionis and S. Sol\`{a}, Phys. Rev. D
\textbf{80} 3511 (2009)

\bibitem {Ar4}J.A.S. Lima, F.E. Silva and R.C. Santos, Class. Quantum Grav.
\textbf{25} 205006 (2008)

\bibitem {Ar5}A.Y. Kamenshchik, U. Moschella and V. Pasquier, Phys. Lett. B
\textbf{511}, 265 (2001)

\bibitem {Ar6}S. Pan, J. de Haro, A. Paliathanasis and R.J. Slagter, Mon. Not.
R. astron. Soc. \textbf{460,} 1445 (2016)

\bibitem {Ar7}A. Paliathanasis, S. Pan and S. Pramanik, Class. Quant. Grav.
32, 245006 (2015)

\bibitem {Ar8}R.C. Nunes and S. Pan, Mon. Not. R. astron. Soc. \textbf{459,}
673 (2016)

\bibitem {ns0}R. de Ritis, G. Marmo, G. Platania, C. Rubano, P. Scudellaro and
C. Stornaiolo, Phys. Rev. D. \textbf{42,} 1091 (1990)

\bibitem {ns2}Y. Kucukakca, U. Camci and I. Semiz, Gen. Relativ. Gravit.
\textbf{44,} 1893 (2012)

\bibitem {ns2a}S. Capozziello, S. Nesseris and L. Perivolaropoulos, JCAP
\textbf{0712,} 009 (2007)

\bibitem {ns3}Y. Kucukakca, Eur.Phys.J. C \textbf{74,} 3086 (2014)

\bibitem {ns4}B. Vakili and F. Khazaie, Class.\ Quantum Gravit. \textbf{29,}
035015 (2012)

\bibitem {ns4a}B. Vakili, Phys. Lett. B \textbf{664,} 16 (2008)

\bibitem {AnScT}A. Paliathanasis, M.\ Tsamparlis, S. Basilakos and S.
Capozziello, Phys. Rev. D \textbf{86,} 063532 (2014)

\bibitem {ns5}R.C. de Souza, R.\ Adre and G.M. Kremer, Phys. Rev. D
\textbf{87,} 083510 (2013)

\bibitem {ns7}P.A.\ Terzis, N.\ Dimakis and T. Christodoulakis, Phys.\ Rev.\ D
\textbf{90,} 123543 (2014)

\bibitem {anb}A. Paliathanasis, M. Tsamparlis, S. Basilakos and J.D. Barrow,
Phys. Rev. D \textbf{93,} 043528 (2016)

\bibitem {belin01}J. A. Belinch\'{o}n, T. Harko and M.K. Mak, Astrophys Space
Sci. \textbf{361,} 52 (2016)

\bibitem {hao}H. Wei, X.J Guo and L.F. Wang, Phys. Lett. B \textbf{707,} 298 (2012)

\bibitem {biswas}B. Modak, S. Kamilya and S. Biswas, Gen. Relativ. Gravit.
\textbf{32,} 1615 (2000)

\bibitem {kt1}A. Paliathanasis, M. Tsamparlis and S. Basilakos, Phys.\ Rev. D
\textbf{90,} 103524 (2014)

\bibitem {kt2}A. Paliathanasis, Class. Quantum Gravit. \textbf{33,} 075012 (2016)

\bibitem {ref01}P. Fr\'{e}, A. Sagnotti and A.S. Sorin, Nucl. Phys. B
\textbf{877,} 1028 (2013)

\bibitem {ref02}V.V.\ Sokolov and A.S. Sorin, Integrable Cosmological
Potentials, [arXiv:1608.08511]

\bibitem {ref03}A.Yu. Kamenshchik, E.O. Pozdeeva, A. Tronconi, G. Venturi and
S.Yu. Vernov, Class. Quantum Grav. \textbf{31,} 105003 (2014)

\bibitem {ref04}A. Paliathanasis, J.D Barrow and P.G.L. Leach, Phys.\ Rev. D
\textbf{94, }023525 (2016)

\bibitem {ref06}N.\ Dimakis, A. Karagiorgos, A. Zampeli, A. Paliathanasis,\ T.
Christodoulakis and P.A.\ Terzis, Phys.\ Rev. D \textbf{93,} 123518 (2016)

\bibitem {ref07}A. Zampeli, T. Pailas, P.A. Terzis and T. Christodoulakis,
JCAP \textbf{05,} 066 (2016)

\bibitem {ref08}H. Suzuki,\ E. Takasugi and Y.\ Takayama, Mod. Phys. Lett.
\textbf{11,} 1281 (1996)

\bibitem {ref09}P. Parsons and J.D. Barrow, Class. Quantum Gravit.
\textbf{12}, 1715 (1995)

\bibitem {ref10}J.D. Barrow and P. Parsons, Phys. Rev. D \textbf{52,} 5576 (1995)

\bibitem {ref11}J.D. Barrow, Phys.\ Rev. D \textbf{48,} 1585 (1993)

\bibitem {ref12}E.O. Kahya, B. Pourhassan and S. Uraz, Phys. Rev. D
\textbf{92,} 103511 (2015)

\bibitem {refAA1}A.A. Coley, J. Ib\'{a}\~{n}ez and I. Olasagasti, Phys. Lett.
A \textbf{250,} 75 (1998)

\bibitem {refAA2}A. Billyard, A. Coley and J. Ib\'{a}\~{n}ez, Phys. Rev. D
\textbf{59,} 023507 (1998)

\bibitem {Jord}P. Jordan, Nature 164, 637 (1937) and \textit{Schwerkraft und
Weltfall,} 2nd ed. (Vieweg und Sohn, Braunschweig, 1955)

\bibitem {BD}C.H. Brans and R.H Dicke, Phys. Rev. \textbf{124}, 925 (1965);
L.E. Gurevich, A.M. Finkelstein and V.A. Ruban, Astrophys. Space Sci.
\textbf{22}, 231 (1973); D. J. Holden, D. Wands, Class.Quant.Grav.,
\textbf{15}, 3271 (1998); J.D. Barrow, Mon. Not. R. Astr. Soc., \textbf{282},
1397 (1996)

\bibitem {ssen}S. Sen and A.A. Sen, Phys. Rev. D \textbf{63,} 124006 (2001)

\bibitem {Boi}B. Boisseau, G. G. Esposito-Farese, D. Polarski and A. A.
Starobisky, Phys. Lett., \textbf{85} 2236 (2000); G. Esposito-Farese and D.
Polarski, Phys. Rev. D. \textbf{63} 063504 (2001); D.E Torres, Phys. Rev. D.
\textbf{66} 043522 (2002); T. Clifton, J.D. Barrow and R. Scherrer, Phys. Rev.
D \textbf{71}, 123526 (2005)

\bibitem {Clif12}T. Clifton, P. G. Ferreira, A. Padilla and C. Skordis, Phys.
Rep., \textbf{513}, 1 (2012)

\bibitem {DEAmendola}L. Amendola and S. Tsujikawa, Dark Energy Theory and
Observations, Cambridge University Press, Cambridge UK, (2010)

\bibitem {omegaBDGR}V. Faraoni, Phys. Rev. D\textbf{59}, 084021, (1999)

\bibitem {BER}B. Bertotti, L. Iess and P. Tortora, Nature \textbf{425}, 374 (2003)

\bibitem {BDconstraints}A. Scharer, R. Angelil, R. Bondarescu, P. Jetzer,A.
Lundgren, Phys. Rev. D., \textbf{90}, 123005 (2014)

\bibitem {Lid}A. R. Liddle, A. Mazumdar and J. D. Barrow, Phys. Rev. D 58,
027302 (1998)

\bibitem {Chen}X. Chen and M. Kamionkowski, Phys. Rev. D 60, 104036 (1999).

\bibitem {Sotiriou}T.P. Sotiriou and V. Faraoni Rev. Mod. Phys. \textbf{82,}
451 (2010).

\bibitem {string}C.G. Callan, D. Friedan, E.J. Martinez, and M.J. Perry, Nucl.
Phys. B \textbf{262}, 593 (1985); E.S. Fradkin and A.A. Tseytlin, Nucl. Phys.
B \textbf{261}, 1 (1985).

\bibitem {cho}Y. M. Cho, Phys. Rev. Lett \textbf{68}, 3133 (1992)

\bibitem {faraonibook}V. Faraoni, \textit{Cosmology in Scalar-Tensor Gravity},
Fundamental Theories of Physics vol. 139, (Kluwer Academic Press: Netherlands, 2004)

\bibitem {bmaeda}J.D. Barrow and K-i Maeda, Nucl. Phys. B \textbf{341}, 294 (1990)

\bibitem {bdpaper}C.H. Brans and R.H. Dicke, Phys. Rev. \textbf{124,} 925 (1961)

\bibitem {Hanlon}J. O'Hanlon, Phys. Rev. Lett. \textbf{29,} 137 (1972)

\bibitem {sarlet}W. Sarlet and F. Cantrijn, SIAM Review \textbf{23,} 467 (1981)

\bibitem {Kalotas}T.M. Kalotas and B.G. Wybourne, J. Phys. A \textbf{15} 2077 (1982)

\bibitem {dim1}P.A.\ Terzis, N. Dimakis, T. Christodoulakis, A. Paliathanasis
and M.\ Tsamparlis, J.\ Geom. Phys. \textbf{101,} 52 (2016)

\bibitem {Aermakov}M. Tsamparlis and A. Paliathanasis, J. Phys. A: Math.
Theor. \textbf{45,} 275201 (2012)

\bibitem {dask}C. Daskaloyannis and K. Ypsilantis, J. Math. Phys. \textbf{47,}
042904 (2006)

\bibitem {ref1}O. Hrycyna and M. Szydlowski, JCAP \textbf{12,} 016 (2013)

\bibitem {ref2}R. Garcia-Salcedo, T. Gonzalez and I. Quiros, Phys. Rev. D
\textbf{92,} 124056 (2015)

\bibitem {ref3}A. Cid, G. Leon and Y. Leyva, JCAP \textbf{16}, 027 (2016)

\bibitem {ref4}G.\ Leon, Class. Quant. Grav. \textbf{26,} 035008 (2009)

\bibitem {bshaw}J.D. Barrow and D.J. Shaw, Class Quantum Gravity \textbf{25},
085012 (2008)


\end{thebibliography}
\end{document}